\documentclass[sn-mathphys,Numbered]{sn-jnl}
\usepackage{graphicx}%
\usepackage{multirow}%
\usepackage{amsmath,amssymb,amsfonts}%
\usepackage{amsthm}%
\usepackage{mathrsfs}%
\usepackage[title]{appendix}%
\usepackage{xcolor}%
\usepackage{textcomp}%
\usepackage{manyfoot}%
\usepackage{booktabs}%
\usepackage{algorithm}%
\usepackage{algorithmicx}%
\usepackage{algpseudocode}%
\usepackage{listings}%
\usepackage{caption}
\usepackage{hyperref}
\usepackage{setspace} 
\usepackage{geometry}
\hypersetup{
	colorlinks=true,
	linkcolor=black,
}
\usepackage{siunitx}
\usepackage{lineno}
\usepackage{setspace}
\makeatletter

\makeatletter
\usepackage[skip=15pt]{caption}
\usepackage{siunitx}%
\usepackage{indentfirst}%
\usepackage[version=4]{mhchem}%
\usepackage{natbib}
\setcitestyle{super,open={},close={}}
\makeatletter
\renewcommand{\@biblabel}[1]{#1.}

\makeatother
\theoremstyle{thmstyleone}%
\theoremstyle{thmstyletwo}%
\theoremstyle{thmstylethree}%
\begin{document}
	\renewcommand{\thelinenumber}{\normalfont\fontsize{10}{12}\selectfont\arabic{linenumber}} 
\title[Article Title]{Optimal Frequency in Second Messenger Signaling: Quantifying cAMP Information Transmission in Bacteria  }

\author[1,2]{\fnm{Jiarui Xiong} \sur{}}
\equalcont{Jiarui Xiong, Liang Wang and Jialun Lin contributed equally to this work.}
\author[3]{\fnm{Liang Wang} \sur{}}
\equalcont{Jiarui Xiong, Liang Wang and Jialun Lin contributed equally to this work.}
\author[2]{\fnm{Jialun Lin} \sur{}}
\equalcont{Jiarui Xiong, Liang Wang and Jialun Lin contributed equally to this work. \newpage}

\newpage
\author[2,4]{\fnm{Lei Ni} \sur{}}
\author[2,4]{\fnm{Rongrong Zhang} \sur{}}
\author[6,7]{\fnm{Shuai Yang} \sur{}}
\author[2,4]{\fnm{Yajia Huang} \sur{}}
\author*[3,5]{\fnm{Jun Chu} \sur{}}
\author*[2,4]{\fnm{Fan Jin} \sur{}}\email{Fan Jin, fan.jin@siat.ac.cn; Jun Chu, jun.chu@siat.ac.cn}

\affil[1]{\orgdiv{Hefei National Research Center for Physical Sciences at the Microscale, Department of Polymer Science and Engineering}, \orgname{University of Science and Technology of China}, \orgaddress{\street{} \city{Heifei}, \postcode{230026}, \state{} \country{China}}}
\affil[2]{\orgdiv{CAS Key Laboratory of Quantitative Engineering Biology, Shenzhen Institute of Synthetic Biology, Shenzhen Institute of Advanced Technology}, \orgname{Chinese Academy of Sciences}, \orgaddress{\street{} \city{Shenzhen}, \postcode{518055}, \state{} \country{China}}}
\affil[3]{\orgdiv{Research Center for Biomedical Optics and Molecular Imaging, Shenzhen Key Laboratory for Molecular Imaging, Guangdong Provincial Key Laboratory of Biomedical Optical Imaging Technology, Shenzhen Institute of Advanced Technology}, \orgname{Chinese Academy of Sciences}, \orgaddress{\street{} \city{Shenzhen}, \postcode{518055}, \state{} \country{China}}}
\affil[4]{\orgdiv{Shenzhen Synthetic Biology Infrastructure, Shenzhen Institute of Synthetic Biology, Shenzhen Institutes of Advanced Technology}, \orgname{Chinese Academy of Sciences}, \orgaddress{\street{} \city{Shenzhen}, \postcode{518055}, \state{} \country{China}}}
\affil[5]{\orgdiv{Key Laboratory of Biomedical Imaging, Shenzhen Institute of Advanced Technology}, \orgname{Chinese Academy of Sciences}, \orgaddress{\street{} \city{Shenzhen}, \postcode{518055}, \state{} \country{China}}}
\affil[6]{\orgdiv{National Science Library (Chengdu)}, \orgname{Chinese Academy of Sciences}, \orgaddress{\street{} \city{Chengdu}, \postcode{610299}, \state{} \country{China}}}
\affil[7]{\orgdiv{Department of Information Resources Management, School of Economics and Management}, \orgname{ University of Chinese Academy of Sciences}, \orgaddress{\street{} \city{Beijing}, \postcode{100049}, \state{} \country{China}}}
\abstract{\fontsize{10pt}{18pt}\selectfont Bacterial second messengers are crucial for transmitting environmental information to cellular responses. However, quantifying their information transmission capacity remains challenging. Here, we engineer an isolated cAMP signaling channel in \textit{Pseudomonas aeruginosa} using targeted gene knockouts, optogenetics, and a fluorescent cAMP probe. This design allows precise optical control and real-time monitoring of cAMP dynamics. By integrating experimental data with information theory, we reveal an optimal frequency for light-mediated cAMP signaling that maximizes information transmission, reaching about 40 bits/h. This rate correlates strongly with cAMP degradation kinetics and employs a two-state encoding scheme. Our findings suggest a mechanism for fine-tuned regulation of multiple genes through temporal encoding of second messenger signals, providing new insights into bacterial adaptation strategies. This approach offers a framework for quantifying information processing in cellular signaling systems.  }

\maketitle
\section*{Main}
In the face of fluctuating external environments, microorganisms must acquire environmental information, make decisions, and adapt to their surroundings by adjusting their behaviors\cite{perkins2009strategies,balazsi2011cellular,bowsher2014environmental,guillemin2020noise}.  In this process, second messengers play a crucial role as mediators, regulating downstream targets in response to external signals (primary signals)\cite{gomelsky2011camp,newton2016second}. The transmission of environmental signals through second messengers to achieve downstream target regulation can be likened to a bow-tie structure in biological systems\cite{friedlander2015evolution}. This structure is characterized by a multi-layered network with a significantly lower number of intermediate components (referred to as the ``waist") compared to the input and output layers. In this analogy, second messengers occupy the mid-layer, which is vital for the transmission of environmental signal information to downstream targets\cite{friedlander2015evolution,newton2016second}.

While it is expected that the versatile regulatory capacity of second messengers relies on their ability to efficiently transmit and encode information, there is a lack of quantitative exploration regarding the upper limits of information transmission mediated by these messengers. Second messengers are known to regulate multiple downstream genes through transcription factors\cite{newton2016second} to achieve specific functions, for instance, through coupled oscillations, cAMP achieves multigenerational memory and adaptive adhesion in bacterial biofilm communities, leading to the emergence of a surface-sentient state\cite{lee2018multigenerational},  but the precise mechanisms of regulation often remain unclear. It is hypothesized that spatial and temporal encoding of second messenger signals may allow for more nuanced control of downstream gene expression beyond simple on/off regulation\cite{schwaller2010cytosolic}. To quantitatively describe the ability of second messengers to transmit and encode environmental information, potentially enabling fine-tuned regulation of multiple genes, we introduce the fundamental principles of information theory\cite{shannon1948mathematical,cover2006elements}. Information theory, a scientific theory established by Shannon, provides a mathematical framework for quantifying data compression and transmission. Shannon's foundational contributions in information theory established that channel capacity represents the ultimate rate of reliable data transmission in communication.

Pioneering studies that applied information theory to biological research have revealed the remarkable efficacy of neuronal systems in information processing and transmission \cite{strong1998entropy,borst1999information,quian2009extracting}. Subsequent investigations have extended information theoretic approaches to intracellular signaling processes, with a focus on quantifying the amount of information reliably transmitted between senders and receivers within the systems under study\cite{tkavcik2008information,cheong2011information,uda2013robustness,selimkhanov2014accurate,uda2020application}.

However, due to the complexity involved in acquiring and processing experimental data, there is a significant dearth of quantitative investigations into the intracellular information transmission capacity of bacterial second messengers. To address this gap, we developed a novel experimental system that combines optogenetic tools and molecular probes to explore the potential of the second messenger cAMP in \textit{Pseudomonas aeruginosa} for transmitting light frequency information. Our approach is innovative in two key aspects: First, we engineered bacterial strains to sever the natural cAMP ``input" pathways and replaced them with light-controlled cAMP production via optogenetics, allowing for precise, programmable control of cAMP levels. Second, we disrupted the downstream transcriptional responses, enabling us to study cAMP as an isolated information channel without the confounding effects of gene expression changes. To achieve both ``writing" and ``reading" of cAMP signals, we carefully designed the system to decouple the wavelengths used for optogenetic activation and fluorescent probe excitation.

By integrating principles from information theory with this experimental setup, we conducted a comprehensive quantitative analysis to characterize the upper limit of the information transmission rate for the cAMP signaling channel within individual bacterial cells in response to light frequencies. Our experimental findings revealed several key insights: we identified an optimal frequency for information transmission, quantified the maximum information transfer rate, and uncovered evidence of a binary encoding scheme used by bacteria at this optimal frequency. These results not only demonstrate a remarkable efficiency in cAMP-mediated information transmission but also provide compelling theoretical support for the potential of cAMP to enable fine-tuned regulation of multiple genes through temporal encoding of signals. Our findings open new avenues for understanding how bacteria might achieve nuanced control of gene expression through second messenger dynamics, bridging the gap between information theory and bacterial regulatory mechanisms
\subsection*{Engineering an Isolated cAMP Signaling Channel with Precise Optical Control and Real-time Monitoring }
Our study aimed to investigate the information transmission capacity of the cAMP signaling pathway in \textit{P. aeruginosa}, a bacterium renowned for its adaptability and involvement in various cellular processes\cite{strateva2009pseudomonas}. To isolate and quantify the information transmission capabilities of cAMP as a signaling molecule, we needed to create a simplified, controllable system that would allow us to study the cAMP channel in isolation from its complex regulatory network. To achieve this, we first identified key proteins in the cAMP signaling pathway of \textit{P. aeruginosa} (Fig. 1a). Adenylyl cyclases CyaA and CyaB\cite{topal2012crystal}, catalyze the synthesis of cAMP from ATP, while the Vfr protein acts as a cAMP-dependent transcription factor\cite{fuchs2010pseudomonas}, transmitting the signal to downstream targets. These components form a bow-tie structure in the signaling pathway, with cAMP at the center.

Our experimental strategy involved three crucial steps to transform this complex network into an isolated channel: (1) Input isolation: We performed gene knockouts of \textit{cyaA} and \textit{cyaB} to eliminate the natural cAMP production in response to environmental signals. This disrupted the connection between the input layer and the cAMP-centered mid-layer. (2) Output isolation: We knocked out the \textit{vfr} gene to sever the link between cAMP and downstream transcriptional responses. This step isolated cAMP from its natural output layer, preventing any feedback or complex gene regulation effects. (3) Controlled input implementation: To precisely control cAMP production, we introduced the optogenetic tool bPAC\cite{stierl2011light} into the genomes of our knockout strains. bPAC, which exhibits adenylate cyclase activity upon blue light stimulation (470 nm), allowed us to use light signals as a programmable input to regulate cAMP synthesis. This carefully designed system enabled us to ``write" cAMP signals into the cell with high temporal precision using light pulses. By eliminating natural inputs and outputs, we created an isolated cAMP channel, allowing us to study its information transmission properties without the confounding effects of the broader regulatory network.

To complete our isolated cAMP channel system, we needed a method to rapidly and accurately ``read'' cAMP signals in real-time. For this purpose, we utilized an optimized version of the red fluorescent cAMP probe, PinkFlamindo2, which we developed based on the original PinkFlamindo\cite{harada2017red} design (Fig. 1b). Our goal was to create a probe with enhanced sensitivity that could more precisely detect cAMP within the physiological concentration range of bacteria. PinkFlamindo2 (abbreviated as PF2) underwent minor amino acid mutations, resulting in significantly improved performance. Based on the dose-response curve (Fig. 1c), PF2 exhibited a lower dissociation constant ($K_d$) of \SI{1}{\micro M}, compared to the original PF2's $K_d$ of \SI{7.2}{\micro M}. This lower $K_d$ value is closer to the physiological cAMP concentration range in bacteria\cite{fulcher2010pseudomonas}, enabling more sensitive signal readouts across a broader range of cAMP levels. Upon binding with cAMP, PF2 exhibits an immediate increase in fluorescence intensity, allowing for real-time monitoring of cAMP fluctuations. A critical aspect of our experimental design was ensuring that the ``writing'' and ``reading'' of cAMP signals could occur simultaneously without interference. We achieved this through the careful selection and optimization of our optogenetic tools. Specifically, bPAC, our ``writing'' tool, shows a maximum absorption peak at 453 nm, while PinkFlamindo2, our ``reading'' tool, has an excitation peak at 570 nm (Fig. 1d). This spectral separation ensures that the activation of bPAC (using 470 nm light) does not interfere with the excitation and measurement of PinkFlamindo2 fluorescence (using 561 nm light). 

In our experiments, we successfully demonstrated the simultaneous application of bPAC and PF2 in our engineered bacterial system. We observed that bacterial strains with low cAMP concentrations exhibited lower fluorescence intensity compared to strains with high cAMP concentrations (Supplementary Fig. S1), confirming the probe's sensitivity to cAMP levels. Importantly, strains carrying bPAC showed stable fluorescence intensity under dark conditions, with significant increases only after light stimulation. This observation further validates the decoupling between the excitation of the cAMP probe and the activation of bPAC (Fig. 1e). Moreover, the fluorescence response of PinkFlamindo2 demonstrated excellent consistency with the opening and closing of the stimulating light source, showcasing the real-time and reversible nature of the probe's response (Fig. 1e).

This carefully designed and optimized system enables us to study the information transmission properties of the cAMP channel with unprecedented temporal resolution and accuracy, paving the way for quantitative analysis of its signaling dynamics and capacity. The successful demonstration of decoupled, real-time control and monitoring of cAMP levels provides a solid foundation for our subsequent investigations into the information transmission capabilities of this second messenger system.
\subsection*{Experimental Observation of Signal and Noise in the cAMP Signaling Channel within Single Bacteria}
For the purpose of experimental convenience and efficiency, we designed a customized multichannel apparatus specifically compatible with the microscope used in our experiments (Fig. 2a, Supplementary Fig. S2, Methods). An LED light source was positioned above the multichannel plate, and the bottom surface of the plate was directly affixed to a 100x oil immersion objective lens. Each channel of the plate could be independently subjected to programmed illumination conditions controlled through a computer interface connected to the microscope control system. The delivery of light stimulation and the acquisition of optical data from the samples were precisely coordinated using a MATLAB script running on the connected PC, allowing for synchronized light delivery and data acquisition across multiple sites using the microscope. As mentioned previously, the wavelengths of the activating light (470 nm) for bPAC and the excitation light (561 nm) for the cAMP probe are distinct. Inspired by theoretical analyses quantifying information transmission rates in response to temporal signal dynamics in the \textit{E. coli} chemotaxis system\cite{tostevin2009mutual}, as well as studies of frequency modulation of transcriptional factor activity in yeast\cite{hansen2015limits}, to facilitate experimental manipulation, we chose a square wave optical signal as our input signal. As shown in Fig. 2a, the cAMP probe within individual bacterial cells exhibited alternating bright and dim states corresponding to the opening and closing of the external light source.

In addition to the experimental procedures, we conducted a comprehensive theoretical analysis of the response characteristics associated with square wave input signals. In a square wave signal with a period of $T$, half of the time is allocated for the presence of external input light, while the other half is in the dark state. Initially, we assumed immediate deactivation of bPAC upon darkness (referred to as Analytical Model 1 to distinguish it from scenarios where bPAC deactivates at a rate post-illumination; see SI section 1.1). Under this assumption, the dynamical behavior of cAMP is governed by the following set of differential equations:
\begin{equation}
	\frac{\text{d[cAMP]}}{\text{d}t} = \begin{cases}
		k_0\text{[bPAC*]}-\gamma \text{[cAMP]}, & \text{if light} , \\
		-\gamma \text{[cAMP]}, & \text{if dark} .
	\end{cases}
	\label{eqtheory1}
\end{equation}
In the equation, [cAMP] represents the concentration of cAMP, [bPAC*] represents the concentration of activated bPAC, and the rate constants for cAMP synthesis by activated bPAC and cAMP degradation by CpdA\cite{fuchs2010v} are denoted by $k_0$ and $\gamma$, respectively. Let $x$ denote the amount of cAMP, and set $k_0[\text{bPAC}^*]=k$. The theoretical analysis of this equation set (see SI section 1.1) reveals that after some time, the oscillation of cAMP will exhibit a periodic pattern, where the peak and trough values during each cycle are denoted as $x_h$ and $x_l$, respectively. The expressions can be derived as:
\begin{equation}
 \begin{cases}
	x_{h}=\frac{k}{\gamma}\frac{1-e^{-\gamma\frac{T}{2}}}{1-e^{-\gamma T}}\\
	x_{l}=x_{h}e^{-\gamma \frac{T}{2}}\
 \end{cases}
 \label{eqhl}
\end{equation}
Here, $T$ represents the period of the input signal. To characterize the information relayed through modulated oscillatory changes, we defined the signal $S$ as the difference between the peak and trough amounts. Based on the analytical formulations, the theoretically obtained expression for the signal is:
\begin{equation}
	S=x_{h}-x_{l}=\frac{k}{\gamma}\frac{(1-e^{-\frac{\gamma T}{2}})^2}{(1-e^{-\gamma T})}
	\label{eqsignal}
\end{equation}
Furthermore, based on the outcome expressed in Equation (\ref{eqhl}), we can deduce the relationship between the amount of cAMP fluctuation over time under cyclic perturbations:
\begin{equation}
	x(t) =
	\begin{cases}
		\dfrac{{\gamma x_{l}-k}}{\gamma}e^{-\gamma t}+\dfrac{k}{\gamma},  & 0\le t\le \frac{T}{2} \\
		x_{h}e^{-\gamma (t-\frac{T}{2})},  & \frac{T}{2}< t\le T 
	\end{cases}
	\label{eqxcycle}
\end{equation}

By means of our experimental manipulation, we were able to record the variations of the fluorescent probe within individual bacterial cells under the input of square wave signals. The fluorescence intensity was subsequently converted into cAMP concentration (Methods). Subsequent data calibration (Methods, Supplementary Fig. S3) enabled a comparative investigation of the oscillatory dynamics of intracellular cAMP levels in bacteria under different periodic conditions. As depicted in Fig. 2b, we examined cAMP fluctuations within single bacterial cells under input conditions with periods of 840 s (low frequency, SI movie 1), 480 s (medium frequency), and 120 s (high frequency) This analysis not only revealed variations in cAMP concentration among bacteria but also, by scrutinizing the changes in the mean values depicted by the solid red line, unequivocally demonstrated that the fluctuations of cAMP were significantly more pronounced during low-frequency input conditions. Conversely, under high-frequency input conditions, it became exceedingly challenging to discern whether the observed cAMP fluctuations stemmed from the input or were induced by intrinsic noise. Moreover, by incorporating the Chemical Reaction Network (CRN) model, we performed fitting of the single-cell data and mean value data to estimate the initial concentrations of CpdA and bPAC in the open model (Supplementary Fig. S4). This approach enabled the determination of two key parameters, $k$ and $\gamma$, as outlined in Equation (\ref{eqsignal}). Consequently, we obtained simulated and theoretical results for periodic fluctuations of cAMP over time after reaching equilibrium. Fig. 2b exhibits the comparison between the experimental mean values (red dots), the analytical oscillations (dark green solid line), and the simulated oscillations (red solid line). It is worth noting that the analytical oscillations are described by Equation (\ref{eqxcycle}). Notably, both the model and the analytical solutions exhibited excellent correspondence with the experimental data.

Considering the inherent variability in CpdA concentration among individual bacteria, we utilized a dimensionless frequency parameter, denoted as $f=\frac{1}{\gamma T}$, to represent the input frequency. This parameter represents the ratio of the input signal frequency to the apparent cAMP hydrolysis rate within the bacteria, allowing for a non-dimensional analysis of frequency response across different bacterial contexts. According to Equation (\ref{eqsignal}), we determine that the maximum value of the signal, $S_{\text{max}}$, is equal to $\frac{k}{\gamma}$. We utilize this value as a normalization criterion to scale the signal values. The relationship between the normalized signal value and the normalized frequency is illustrated in Fig. 2c. In the low-frequency range, both the theoretical analysis (dark green solid line) and the simulation results (red solid line) exhibit close agreement. However, discrepancies emerge in the high-frequency range. This discrepancy arises from the assumption made in our theoretical analysis, where bPAC is assumed to instantaneously deactivate. In contrast, the model used for simulation incorporates a certain deactivation rate for bPAC, which is more consistent with reported literature. In the low-frequency range, the deactivation time is considerably shorter compared to the experimental period, resulting in minimal impact on the analysis. Consequently, the theoretical analysis aligns well with the simulation results in the low-frequency region. As the fitting is conducted using the model for the experimental data, the experimental data points (red dots) demonstrate closer alignment with the corrected theoretical analysis (light green solid line, analytical model 2. SI section 1.2; Supplementary Fig. S5). Nevertheless, all analyses indicate that as the frequency increases, the signal attenuates, reflecting the low-pass characteristics of the cAMP system. In the frequency range where $f\gg 1$, the signal decays following an inverse relationship with the frequency, consistent with the theoretical analysis (SI section 1). 

As widely acknowledged, noise is an intrinsic attribute of biological systems\cite{simpson2009noise,guillemin2020noise}, and the response of cAMP molecules to light frequencies is no exception. In our study, we conducted a detailed analysis of the noise present in this process at the molecular level. To accomplish this, we utilized the CRN model for data fitting. By comparing the experimental data, which exhibited stochastic characteristics, with the ideal curve of the CRN model, which displayed deterministic characteristics, we quantified the noise, represented as $\sigma$, in the cAMP levels within individual bacterial cells (Fig. 2d). By drawing an analogy between the experimental synthesis and degradation of cAMP and classical birth-death process models\cite{allen1990probability}, we can deduce that the number of cAMP molecules follows a Poisson distribution (Supplementary Fig. S6a; SI section 2). Specifically, the average number of cAMP molecules, denoted as $N$, during periodic oscillation is numerically equal to the variance of the distribution, which is the square of the noise, i.e., $N = \sigma^2$. Furthermore, through the analysis of Equation (\ref{eqxcycle}), it is established that $N = \frac{\int_{0}^{T} x(T) \text{d}t}{T} = \frac{k}{2\gamma}$. Thus, theoretically, the magnitude of the noise can be related to the key parameters $k$ and $\gamma$. It is crucial to emphasize that when utilizing these formulas to address noise-related issues, it is necessary to convert the relevant concentration units into molecule counts for accurate analysis. This conversion process necessitates estimating the volume of individual bacteria (Methods, Supplementary Fig. S6b). In our actual experimental observations, we have observed that the squared magnitude of the noise exceeds the average number of cAMP molecules. This observation can be attributed to various factors, including limitations associated with experimental observations and data processing. However, a notable positive correlation exists between the noise and the number of cAMP molecules, suggesting that an increase in the number of cAMP molecules is accompanied by an increase in the magnitude of the noise. Importantly, the aforementioned noise refers to absolute noise, while relative noise is decreasing. It is noteworthy that both the noise and the number of cAMP molecules remain independent of the input frequency (Supplementary Fig. S6c-d).

Once the signal magnitude and noise characteristics of the cAMP channel have been determined, the investigation of the signal-to-noise ratio (SNR) becomes paramount. The SNR concept is widely employed in communication systems\cite{cover2006elements}, where a higher SNR signifies reduced interference from noise during the communication process. Building upon the preceding discussion regarding signal and noise, we define the SNR in the molecular communication of cAMP as the square of the ratio between the signal and the noise, expressed as $\text{SNR} = \frac{S^2}{\sigma^2}$. Consequently, the magnitude of the SNR in the cAMP communication of individual bacteria within the experimental framework can be determined.
From a theoretical perspective, taking into account the aforementioned analysis on signal magnitude and theoretical noise, we can derive the analytical expression for the SNR as follows:
\begin{equation}
	\text{SNR}(f)=\frac{S^2}{\sigma^2}=\frac{2k}{\gamma}\frac{(1-e^{-\gamma\frac{T}{2}})^4}{(1-e^{-\gamma T})^2}=\frac{2k}{\gamma}\frac{(1-e^{-\frac{1}{2f}})^4}{(1-e^{-\frac{1}{f}})^2}
	\label{eqSNRf}
\end{equation}

Hence, for each set of experimental data from individual bacteria, we can normalize it using $\frac{2k}{\gamma}$ as the maximum value. This normalization allows us to establish the relationship between the normalized SNR and the normalized frequency $f$ for the cAMP channel in single bacteria, as illustrated by the red data points in Fig. 2e. Furthermore, leveraging the aforementioned analytical expression, we can plot the theoretical relationship between SNR and $f$, represented by the red line in Fig. 2e. Notably, as $f$ increases, the SNR exhibits a declining trend. As previously discussed, the signal attenuates at higher frequencies, the noise, in contrast, remains independent of the frequency. Drawing from this observation and the previously established attenuation rate of signal magnitude in the high-frequency range, we deduce that the SNR will decay following an $f^{-2}$ trend in the high-frequency range (SI section 3).
The findings concerning the variation of SNR with frequency carry significant implications. Specifically, with high frequency comes low SNR, and with low frequency comes high SNR. In other words, the faster the signal transmission, the lower the accuracy, whereas slower transmission leads to higher accuracy. This discussion naturally leads to a central query: whether an optimal frequency exists that maximizes the information transfer rate within a unit time, consequently optimizing the rate of information transmission. Prior to reaching this optimal frequency, the information transmission rate gradually increases with frequency, but beyond this point, it declines.
\subsection*{Determining the Upper Limit of Information Transmission Rate in Single Bacterial cAMP Channel}
The determination of the SNR forms the basis for elucidating the pivotal inquiry of information transmission rate within the cAMP channel, which constitutes the central objective of this study. Drawing upon Shannon's formulation of channel capacity, we approximate our problem as a Gaussian channel. The upper limit of the information transmission rate in the cAMP channel is denoted as $I$, and its theoretical representation can be expressed as follows:
\begin{equation}
	I(f)=\frac{f\gamma}{2}\text{log}_2\;(1+\text{SNR}(f))
	\label{eqIrate}
\end{equation}
The aforementioned equation calculates the theoretical maximum limit of information that can be transmitted per unit time, expressed in bits/s. In this context, 1 bit represents the ability to distinguish between two possible states of the input based on the output. To facilitate a unified analysis of individual bacteria, which may display varying degradation rates, we introduced a normalization condition based on the information transmission rate achieved at the normalized frequency $f = 1$. This implies that the input frequency is equal to the apparent degradation rate $\gamma$, denoted as $I(f = 1)$. Specifically, $I(f=1)=\frac{\gamma}{2}\text{log}_2\;(1+\text{SNR}(f=1))$. Based on the analysis derived from Equation (\ref{eqSNRf}), it is straightforward to deduce that $\text{SNR}(f=1)=\frac{2k}{\gamma}\frac{(1-e^{-\frac{1}{2}})^4}{(1-e^{-1})^2}=0.24N$. Thus,
\begin{equation}
	I(f=1)=\frac{\gamma}{2}\text{log}_2(1+0.24N)
	\label{eqf1}
\end{equation}
Within this normalization framework, we define the normalized information transmission rate, $I_r(f)$, as the ratio of the information transmission rate at frequency $f$, $I(f)$, to the information transmission rate at $f=1$, $I(f=1)$. Mathematically, it can be expressed as $I_r(f)=\frac{I(f)}{I(f=1)}$.

In the preceding sections, we have empirically established the SNR for the cAMP channel in individual bacteria. By applying Equation (\ref{eqIrate}), we can subsequently ascertain the upper limit of the information transmission rate for each individual bacterium. Additionally, Equation (\ref{eqf1}) can be utilized for the dimensionless normalization process. As a result, the relationship between $I_r$ and $f$ can be obtained, as illustrated in Figure 3a (red dots). In theoretical terms, by utilizing the relationship $N=\sigma^2$, we can derive the analytical expression for the normalized information transmission rate, $I_r$, as follows:
\begin{equation}
	I_r(f)=\frac{I(f)}{I(f=1)}=f\frac{\text{log}_2(1+4N(1-e^{-\frac{1}{2f}})^4/(1-e^{-\frac{1}{f}})^2)}{\text{log}_2(1+0.24N)}
	\label{eqIr}
\end{equation}

The aforementioned equation unveils that, in theory, the upper limit of the relative information transmission rate, $I_r$, can be considered as a function of the normalized frequency, $f$, and the average number of cAMP molecules, $N$. In Figure 3a, the theoretical correlation between $I_r$ and $f$ is graphically represented (red curve) for varying values of $N$ (specifically, $N$ = 10, 100, and 1000). The curves depicted in Figure 3a demonstrate a concave downward shape, providing validation for the existence of the previously hypothesized optimal frequency, denoted as $f^*$, as discussed in the context of the SNR-frequency relationship. At $f^*$, the value of $I_r$ attains its maximum point.

On the other hand, the color shading in Figure 3a corresponds to the average number of cAMP molecules. Darker red indicates higher value of $N$. By examining the figure, it can be observed that curves associated with larger $N$ generally display an rightward trend in the two-dimensional plane. This observation implies a strong correlation between the optimal normalized frequency $f^*$ and $N$.

For each individual bacterium in the experimental observations, the values of $k$ and $\gamma$ can be fitted to determine $N$. This enables the establishment of a theoretical $I_r$-$f$ curve for each bacterium. As described previously, by plotting these curves, the optimal frequency $f^*$ can be identified. Consequently, the relationship between $f^*$ and $N$ can be further elucidated, as depicted in Figure 3b. The results demonstrate a well-fitted mathematical quantitative relationship between $f^*$ and $N$, which can be expressed as follows: $f^*\approx0.255N^{0.5}$. This implies that the magnitude of the optimal normalized frequency increases as the average number of cAMP molecules within one period increases.

Further analysis of this quantitative fitting result reveals an intriguing discovery. By substituting this quantitative result into Equation (\ref{eqSNRf}), it is found that $\text{SNR}(f^*)\approx3.84$ (SI section 3), indicating that at the optimal normalized frequency, the SNR becomes a constant value independent of $f^*$. This result carries an additional implication: at the optimal frequency $f^*$, the information content (distinct from the information rate) also remains constant. Let us denote the number of states corresponding to the information content as $M$; hence, we have:
\begin{equation}
	M=2^{\frac{1}{2}\text{log}_2(1+\text{SNR}(f^*))} \approx 2.2 
	\label{eqM}
\end{equation}
The relationship between $M$ and $f^*$ is illustrated in Figure 3c, unveiling a crucial inference in our study. At the optimal frequency, bacteria employ a binary encoding scheme with two states to achieve the fastest information transmission through the cAMP channel. This result emphasizes a significant finding in our research regarding the utilization of the cAMP signaling pathway by bacteria for efficient optical frequency information transfer.

By combining the findings from Figures 3b and 3c, it can be inferred that within the context of bacterial communication through the cAMP signaling pathway for optical frequency information transfer, the average number of cAMP molecules within one period influences the optimal frequency. Specifically, the optimal frequency increases as the number of molecules increases. However, at the optimal frequency, the maximum transmitted information content remains relatively constant. This observation unveils the adoption of a binary encoding scheme with two states by bacteria when utilizing the cAMP signaling pathway for information transfer at the optimal frequency.

Given that the normalized frequency is determined by the degradation rate $\gamma$ of cAMP, it can be inferred that the maximum information rate, $I_{\text{max}}$, for a bacterium transmitting optical frequency signals is approximately $I_\text{max}\approx0.3\gamma N^{0.5}$. Consequently, by utilizing the degradation rate of cAMP and the average number of cAMP molecules, the upper limit of the information transmission rate at the optimal frequency can be computed for each bacterium. The statistical distribution of this data pertaining to individual bacteria can be found in the supplementary materials (Supplementary Fig. S7). The variation in intracellular biochemical parameters among bacteria results in differences in the upper limit of information transmission rates. This finding holds significant importance as it provides guidance for the rational design of genetic circuits. For example, regulating the intracellular cAMP degradation enzyme offers the possibility of controlling the maximum information rate. This insight paves the way for modulating information transmission capabilities in bacteria through targeted control of specific biochemical parameters, thereby offering valuable implications for applications in synthetic biology. In our experimental system, we have observed that a significant number of single bacterial information transmission rates surpass 0.011 bits/s (refer to Supplementary Figure S7 for the distribution statistics), which is approximately equivalent to 40 bits/h. This noteworthy discovery suggests that during a one-hour interval of bacterial growth and division, the upper limit of cAMP channel information transmission approaches 40 bits. The existence of such a high-efficiency upper limit in information transmission offers theoretical support for the potential of fine-tuned regulation of multiple genes within bacterial systems.
\subsection*{Estimating Channel Capacity by Utilizing Classifiers for Decoding Error Rate}
Recent advancements in mathematical modeling and machine learning techniques have opened up new avenues for quantitative analysis of information transmission in intracellular signaling\cite{granados2018distributed,jetka2019information,tang2021quantifying,cepeda2019estimating,ying2022quantifying}. Building upon these seminal studies, our research endeavors involve quantifying information transmission in the cAMP signaling pathway using Shannon's formula, as depicted in Figure 3. Additionally, we employ chemical reaction modeling and machine learning methods to unravel the mechanisms of cAMP-mediated information transmission in bacteria. Our primary research objective is to develop a predictive model capable of interpreting the light stimulation state based on dynamic changes in cAMP levels within individual cells. To achieve this goal, as illustrated in Figure 4a, we employ the Stochastic Simulation Algorithm (SSA)\cite{gillespie1976general,gillespie1977exact} to generate a simulated dataset, which serves as the training data. Subsequently, we utilize a neural network-based machine learning model to classify the test data, comprising both simulated and experimental data, based on their oscillation characteristics. This classification process decodes the data into two categories: light-illuminated and light-off states. Specifically, we assign the value 1 to the light-illuminated state and 0 to the light-off state. The decoding accuracy achieved on the simulated data subset, as demonstrated in the truth table presented in the figure, exceeds 80\% across all evaluated test periods, considering varying code durations.

In the experimental setup, we implemented periodic random manipulations of the external light source state at defined intervals, which we refer to as code durations. The light source had an equal probability of being turned on or off, encoded as 1 and 0, respectively. Figure 4b illustrates the fluctuations of cAMP in response to random inputs for different code durations of 360 s (SI movie 2), 180 s, and 120 s. The figure demonstrates that the average value, represented by the thick solid red line, exhibits more pronounced fluctuations compared to individual bacterial fluctuations. This observation suggests that decoding based on the average output is more feasible (see SI section 4 and Supplementary Fig. S8 for more details). Furthermore, it is evident that under longer code durations of 180 s and 360 s, corresponding to lower frequencies, the fluctuations of cAMP are more prominent compared to the shorter code duration of 120s. This indicates that it is easier to determine the state of the blue light switch at the input end based on the fluctuation results under these lower frequency conditions. To account for variations in cAMP degradation rates among individual bacteria, similar to previous methodologies, we employed a model-fitting approach to estimate the intracellular concentration of CpdA. This estimation allowed us to calculate the degradation rate constant for cAMP within each bacterium, which was then used to normalize the experimental frequency. 

By integrating the trained machine learning model with cAMP fluctuations as a basis, we decoded the external input signals. The decoded results were compared to the recorded input sequences from the experiments to assess the decoding accuracy. Figure 4c illustrates the decoding accuracy of different bacteria under various conditions. It is evident that as the normalized frequency increases, the decoding accuracy tends to decrease, consistent with the phenomenon of high-frequency signal attenuation observed in periodic signals. Additionally, we conducted model simulations with varying degradation rate constants ($\gamma$) to examine the relationship between decoding accuracy and normalized frequency. The simulation results revealed a slight decrease in decoding accuracy as $\gamma$ increased for the same normalized frequency.

To establish a connection between decoding accuracy and information transmission, we can draw an analogy between our channel and a binary symmetric channel (BSC)\cite{cover2006elements} . In a typical BSC, there is an error probability in decoding a 1 as 0 and vice versa. Based on the results depicted in Figure 4c, we can simulate the decoding accuracy under different $\gamma$ values. Considering the characteristics of our study, we describe the channel properties using the decoding error rate ($p_e$) and treat it as a typical BSC. For such a channel, the channel capacity ($C$) can be calculated using the formula: $C = 1 - H(p_e)$. Here, $H(p_{e})=-p_{e}\text{log}_2p_e-(1-p_e)\text{log}_2(1-p_e)$ represents the entropy calculation. When $p_{e} = 0.5$, the channel capacity becomes 0 bits, indicating the inability to transmit information.
By considering the varying error rates for different code durations ($T_c$), we can determine the upper limit of information transmission rate for each code duration using the following equation:
\begin{equation}
	\frac{C}{T_c}=\frac{1-H(p_e)}{T_c}
\end{equation}

By combining the machine learning results with Equation (10), we can establish the relationship between $\frac{C}{T_c}$ and the frequency $\frac{1}{T_c}$, as illustrated in Figure 4d. From the graphical representation, we observe the presence of an optimal value for $\frac{1}{T_c}$ that maximizes the transmission efficiency, represented by $\frac{C}{T_c}$. As the degradation rate constant ($\gamma$) increases from 0.0012 $\text{s}^{-1}$ to 0.0069 $\text{s}^{-1}$, the transmission rates rises from 0.004 bits/s to 0.011 bits/s.
\section*{Summary and Perspective}
Our study integrates information theory with empirical science to explore the upper limit of information transfer rate facilitated by cAMP, a second messenger molecule, within bacterial cells. While we focused on cAMP in \textit{P. aeruginosa}, the framework we developed has broad applicability to other crucial second messengers across various bacterial species. For instance, our methodology could be readily adapted to study the information transfer dynamics of c-di-GMP, another vital second messenger involved in bacterial biofilm formation\cite{valentini2016biofilms}.

A key finding of our research is the identification of an optimal frequency for information transmission in the cAMP signaling pathway. This non-trivial discovery, reported here for the first time, has profound implications for understanding microbial second messenger regulation. The existence of an optimal frequency suggests that bacteria may have evolved to maximize their information processing capabilities within the constraints of their biochemical machinery. This finding opens new avenues for investigating how bacteria encode and transmit environmental information through their signaling networks.

Our quantification of the upper limit of information transmission rate - nearly 40 bits of information within a 1-hour bacterial division cycle - provides strong theoretical support for the global regulatory capacity of cAMP. This high information capacity aligns with the potential of cAMP to regulate multiple downstream genes, offering a quantitative basis for understanding fine-tuned regulation in bacteria. According to the information processing inequality\cite{cover2006elements}, this channel could theoretically modulate the activity of up to 40 independent downstream genes. However, it's important to note that in reality, gene expression is often interdependent\cite{klosik2017interdependent}, which may affect how this information capacity translates to actual regulatory outcomes.
The variation in cAMP hydrolysis rates among individual bacteria, resulting in diverse information transfer rates, points to an intriguing aspect of bacterial population dynamics. This heterogeneity may represent a bet-hedging strategy\cite{veening2008bistability}, allowing bacterial populations to maintain resilience in fluctuating environments. Our finding that the optimal frequency of information transmission is influenced by cAMP molecule quantity, while maintaining a consistent binary encoding scheme, further underscores the sophistication of bacterial signaling systems.

While our study provides a robust framework for investigating the upper limits of information transmission by second messengers, several questions remain for future exploration. For instance, the extent to which downstream targets utilize the transmitted information and how this compares to other well-studied systems, such as \textit{Escherichia coli} chemotaxis\cite{mattingly2021escherichia}, requires further investigation. Additionally, the complexity introduced by the interdependence of gene expression and the potential interactions between different second messengers in real-world environments presents exciting challenges for future research.

In conclusion, our study represents a significant step forward in applying information theory to bacterial signaling systems. By quantifying the information transmission capacity of the cAMP channel and identifying an optimal signaling frequency, we have provided new insights into how bacteria might achieve fine-tuned regulation of multiple genes through second messenger dynamics. These findings not only enhance our understanding of bacterial adaptation and decision-making processes but also open up new possibilities for synthetic biology applications, where precise control of cellular behavior is desired. As we continue to unravel the complexities of microbial signaling networks, the integration of information theory with molecular biology promises to yield further insights into the remarkable information processing capabilities of these seemingly simple organisms.

\section*{Methods}
\subsection*{Construction of Bacterial Strains}
The bacterial strain utilized in this study was derived through genetic manipulation of \textit{P. aeruginosa}, a well-established model organism in this field. The knockout procedure involved the utilization of the pEX18Gm plasmid to express both upstream and downstream sequences of the target gene\cite{huang2017rapid}, resulting in the successful generation of knockout plasmids, namely pEX18Gm-vfr, pEX18Gm-cyaA, pEX18Gm-cyaB, and pEX18Gm-cpdA. The upstream and downstream sequences, approximately 1000 bp in length, were amplified from the bacterial genome using PCR and subsequently assembled into the linearized pEX18Gm vector through Gibson Assembly. By employing these knockout plasmids, the bacterial strains were effectively engineered to completely lack the specific genes, thus establishing a genetically unaltered background for subsequent experimental investigations.

Before introducing the optogenetic tool bPAC and the molecular probe tool PinkFlamindo2 (PF2) into our host bacterium, we conducted codon optimization\cite{grote2005jcat} to enhance their gene expression levels in PAO1. Subsequently, the bPAC gene was amplified by PCR and ligated into the ctx2 vector, followed by standardized procedures\cite{hoang2000integration,hoang1998broad} for integration into the PAO1 genome with \textit{cyaA}, \textit{cyaB}, and \textit{vfr} genes knocked out. The cAMP molecular probe tool PF2 was incorporated into the pUCP20 plasmid using Gibson Assembly\cite{gibson2009enzymatic}, with the PF2 gene being expressed under the control of the constitutive promoter J23100 on the same plasmid. Afterwards, the plasmid containing the cAMP molecular probe was introduced into the corresponding PAO1 variant bacterial strains through electroporation. The pivotal strain, referred to as the bPAC-PF2 strain, was generated by specifically disrupting the \textit{cyaA}, \textit{cyaB} and \textit{vfr} genes and integrating the bPAC gene into the genome. This strain, which expresses the bPAC gene in the genome and the PF2 reporter on a plasmid, represents the central focus of the study.
\subsection*{Cultivation Protocol for Bacterial Growth}
The experiment commenced by inoculating the frozen experimental bacterial strains, stored at \SI{-80}{\degreeCelsius}, onto LB agar supplemented with \SI{30}{\micro\gram/\milli\liter} gentamicin. The agar plates were then incubated at \SI{37}{\degreeCelsius} for 16 hours to facilitate the revival of the bacteria. Subsequently, the revived bacterial colonies were resuspended to \SI{1}{\milli\liter} of fresh FAB medium containing \SI{30}{\micro\gram/\milli\liter} gentamicin, \SI{1}{\micro M} ferric chloride, and \SI{100}{\milli M} glycerol until the optical density ($\text{OD}_{600}$) reached approximately 1.0. Next, \SI{3}{\micro\liter} of the resuspended bacterial culture was aliquoted and diluted in 1 mL of fresh FAB medium containing the same components. The diluted culture was then incubated at \SI{37}{\degreeCelsius} with continuous shaking at 220 rpm for 12 hours, allowing overnight growth. Following overnight growth, the 1 mL bacterial culture was transferred to a \SI{1.5}{\milli\liter} centrifuge tube and centrifuged at 3300 rpm for 3 minutes. Approximately \SI{900}{\micro\liter} of the supernatant was then aspirated, and the remaining liquid was gently resuspended by pipetting, resulting in a concentrated bacterial sample suitable for subsequent microscopic examination.
\subsection*{Sample Preparation for Microscopic Observation}
Following the culture and harvest of the target bacterial samples, a multi-well agarose gel sandwich assay was employed to immobilize and array the cells for subsequent microscopic observation. Briefly, a dual-layer seven-well plate, composed of top and bottom sections magnetically joined at the periphery, was utilized (Supplementary Fig. S2). The bottom section was equipped with a circular coverslip suitable for microscope stage-mounting, and an intervening circular coverslip was incorporated between the layers. A defined culture medium, consisting of 1.5\% agarose FAB broth, was prepared by boiling for clarification, with the same FAB composition as previously described. Subsequently, \SI{120}{\micro\liter} of the liquified agarose-medium solution, in its sol phase prior to gelling, was pipetted into each well of the plate. Upon cooling and setting at room temperature, the agarose solidified, forming a gel substrate that contacted the intervening coverslip positioned between the plate sections. This arrangement generated a layered agarose scaffold, which served as the substrate arrayed within the multi-well assembly. To immobilize the bacterial samples, the plate layers were separated, and the intervening coverslip was lifted to allow the targeted deposition of \SI{1.5}{\micro\liter} of the prepared bacterial samples onto the exposed agarose substrates. Rejoining the magnetic plate sections embedded the agarose monolayers, along with the affixed cells, into the bottom section for subsequent microscopic examination. Through this multi-layer agarose gel entrapment approach within a partitioned microplate assembly, uniform and high-throughput immobilization of the target bacterial specimens was achieved in a format readily adaptable to microscopic investigation under controlled conditions.
\subsection*{Microscopic Observation Experiment}
Samples were examined using an Olympus IX81 inverted fluorescence microscope equipped with a 100× oil immersion objective. The microscope's LED illumination and functions were computer-controlled using a MATLAB interface. Specifically, the samples were placed on the stage under the 100× oil objective and illuminated from above using an LED lamp. The LED light and image acquisition were controlled through a connected computer running a MATLAB program\cite{fu2023programming}. This setup facilitated computerized fluorescent imaging of the samples at high magnification under oil immersion. For bPAC stimulation, the LED light intensity was set at \SI{30}{\micro\watt/\centi\meter\squared}, and a 300 ms exposure was used for PF2 excitation. The MATLAB program allowed us to control the periodic or random cycling of the LED on-off states, with the sampling start time and intervals set according to experimental requirements to acquire fluorescence images of the probe. The acquired fluorescence images were computationally analyzed using a MATLAB code previously developed in our laboratory\cite{chen2022genome}. The image data were processed through the program, enabling the quantification of fluorescence intensities over time at the resolution of single bacterial cells.
\subsection*{Procedure for Conversion of Fluorescence Intensity to cAMP Concentration and Data Calibration}
The fluorescence intensity of the cAMP-binding probe PF2 exhibits an increase in response to cAMP binding, enabling the characterization of intracellular cAMP concentration. The measured fluorescence ($\Phi$) comprises two components: the fluorescence emitted by free PF2 ($\Phi_f$) and the fluorescence emitted upon the interaction of PF2 with cAMP ($\Phi_c$). By assuming a binding fraction ($z$) that represents the proportion of PF2 bound to cAMP out of the total PF2, and the total concentration of PF2 ($\zeta$), we can describe the fluorescence intensity using the following equation:
\begin{equation}
	\Phi=\Phi_c+\Phi_f=K_c\zeta z+K_f\zeta(1-z)
	\label{eqIntensity}
\end{equation}
here,
$K_c$
represents the luminescence coefficient of the cAMP-probe complex, while
$K_f$
represents the luminescence coefficient of the unbound cAMP probe.
The equation mentioned above incorporates the luminescence coefficient, which can be determined by considering the observed dynamic range of the cAMP probe. Specifically, we can express this coefficient as $E=\frac{\Delta \Phi}{\Phi_0}=\frac{\zeta K_c-\zeta K_f}{\zeta K_f}$, where $\Phi_0$ denotes the baseline fluorescence intensity. Analysis of the results presented in Supplementary Figure 1 (Supplementary Fig. S1) suggests that the numerical value of $E$ is approximately 6. Besides, the Equation (11) establishes a fundamental relationship between fluorescence intensity and the binding fraction. 

Moving forward, it is imperative to elucidate the precise correlation between the binding fraction ($z$) and the concentration of cAMP. The binding process of cAMP to PF2 can be accurately represented by the following reversible chemical reaction equation (Supplementary Table S2):
\begin{equation}
	\ce{cAMP + PF\text{2} <=>[$k_{7f}$][$k_{7r}$] cAMP-PF\text{2}}
\end{equation}
wherein cAMP and PF2 undergo reversible binding to form the cAMP-PF2 complex, characterized by the forward rate constant $k_{7f}$ and the reverse rate constant $k_{7r}$ (Supplementary Table S4). The dynamics of the cAMP-PF2 complex can be mathematically described by the following differential equation:
\begin{equation}
	\frac{\text{d[cAMP-PF2]}}{\text {d}t}=k_{7f}[\text{cAMP}][\text{PF2}]-k_{7r}[\text{cAMP-PF2}]
\end{equation}
This differential equation represents the dynamic equilibrium between the association and dissociation processes of cAMP and PF2, enabling the quantification of binding kinetics by determining the previously defined forward and reverse rate constants. At equilibrium, the rate of formation of the cAMP-PF2 complex is equal to the rate of its dissociation. By solving the aforementioned differential equation, the concentration of the cAMP-PF2 complex at equilibrium can be derived as follows:
\begin{equation}
	[\text{cAMP-PF2}] = \frac{k_{7f}[\text{cAMP}][\text{PF2}]}{k_{7r}[\text{cAMP-PF2}]}
\end{equation}

Based on the previously established definition, the binding fraction, $z$, is defined as the ratio of the concentration of the cAMP-PF2 complex [cAMP-PF2] to the total concentration of the complex and free PF2, [cAMP-PF2] + [PF2]. Additionally, the dissociation constant, $K_d$, is defined as the ratio of the reverse rate constant $K_{7r}$ to the forward rate constant $k_{7f}$, $K_d=\frac{k_{7f}}{k_{7r}}$. Utilizing these relationships, the concentration of cAMP can be determined in terms of the binding fraction $z$ as follows:
\begin{equation}
	x=K_d\frac{z}{1-z}
\end{equation}
By combining Equations (11) and (15), incorporating the defined parameter $E$, and defining the parameters $a=K_f \zeta$, we can establish the relationship between cAMP concentration ($x$) and the observed fluorescence intensity ($\Phi$) as follows:
\begin{equation}
	x=K_d\frac{\Phi -a}{a(E+1)-\Phi}
\end{equation}

In the aforementioned equation, the variable $x$ represents the cAMP concentration, $\Phi$ represents the observed fluorescence intensity of the probe, $K_d$ denotes the dissociation constant between cAMP and the probe, expressed in concentration units, and $E$ represents the dynamic range. The value of parameter $a$ can be estimated based on relevant literature findings. For example, in PAO1 strains with \textit{cyaA} and \textit{cyaB} genes knocked out, the reported cAMP concentration is approximately 20 nM\cite{fulcher2010pseudomonas}. From the fluorescence intensity of this bacterial strain observed in our experiment, a rough estimate for parameter $a$ is approximately 14. It is important to note that the fluorescence intensity can be influenced by various sampling conditions, and recalibration of parameter $a$ is necessary when altering experimental conditions. 

While analyzing the observed fluorescence intensity data from our experiments, we encountered a phenomenon where the converted cAMP concentration exhibited attenuation throughout the microscopy experiments, particularly during periodic illumination tests using the bPAC-PF2 bacterial strain. This attenuation may be attributed to the photobleaching effects caused by frequent sampling, which resulted in a decay in the analyzed cAMP concentration over time. However, according to our theoretical analysis, the cAMP concentration should exhibit stable oscillatory fluctuations under periodic illumination. To improve the fitting of our data to theoretical models, we performed a manual correction on the data. Specifically, we extracted the peak values of the oscillating cAMP signals and applied an exponential decay function to fit the decay rate. This decay rate was then used to correct the original data, accounting for the changing cAMP concentration over time (Supplementary Fig. S3).
\subsection*{Estimation of Cell Volume }
Based on the image analysis, we obtained the following parameters: cell area, width, and length. By referencing existing literature\cite{taniguchi2010quantifying} and considering that each pixel in our observations using a 100x objective lens corresponds to a length of $\frac{\SI{6.5}{\micro m}}{100}$, we estimated the cell volume as $V=0.065 \times 2/3 \times \text{area} \times \text{width}$, with the resulting unit in \SI{}{\micro m^3}.
To convert the concentration of intracellular molecules into the number of molecules, we leverage Avogadro's constant. The distribution of calculated volumes  can be found in Supplementary Information. (Supplementary Fig. S6b).
\backmatter
\section*{Supplementary Information}
Additional results, additional tables and figures can be found in Supplementary Information. 
\section*{Acknowledgements}
Special thanks to Aiguo Xia for guidance and assistance with microscope usage, to Lei Wang for developing the experimental setup and to Shengjie Wan for valuable suggestions during simulations. This work was supported by the National Key Research and Development Program of China (Grant No. 2020YFA0906900  to Fan Jin), the National Natural Science Foundation of China (Grant No. 32101177 to Yajia Huang), Shenzhen Engineering Research Center of Therapeutic Synthetic Microbes (Grant No. XMHT20220104015 to Fan Jin), the National Natural Science Foundation of China (Grant No. 32371431 to Jun Chu) and the Innovation Foundation of National Science Library (Chengdu) (Grant No. E3Z0000903 to Shuai Yang).
\section*{Author Contributions}
We were responsible for this study as follows: Conceptualization, Fan Jin; Development and characterization of the probe, Liang Wang and Jun Chu; Codon optimization, primer design and strain construction: Jiarui Xiong and Yajia Huang; Method of utilizing the seven well plate device and LED lights: Lei Ni and Jiarui Xiong; Exploration of bacterial culture conditions: Jiarui Xiong and Lei Ni; Correction of light intensity power between different holes of LED lights: Rongrong Zhang and Jiarui Xiong; Optimization of lighting conditions: Jiarui Xiong and Rongrong Zhang; Getting data from the periodic light exposure experiment: Jiarui Xiong; Code for analysis of fluorescence intensity in single bacteria: Lei Ni and Shuai Yang; Method of converting fluorescence intensity into cAMP concentration: Fan Jin; Construction of a Chemical Reaction Network model: Jialun Lin; Theoretical analysis of cAMP dynamics: Fan Jin and Jiarui Xiong; Correction of experimental data: Jiarui Xiong; Fitting of experimental data: Jialun Lin; Analysis of theoretical noise: Fan Jin and Jiarui Xiong; Method for converting cAMP concentration to molecule count: Lei Ni and Jiarui Xiong; Analysis of experimental noise: Jialun Lin; Theoretical analysis of SNR: Fan Jin and Jiarui Xiong; Proposal of the optimal frequency: Fan Jin; Theoretical analysis of the upper limit of information transmission rate: Fan Jin and Jiarui Xiong; Proposal of the encoding scheme: Fan Jin; Proposal of the random input method: Fan Jin; Acquisition of experimental data on random light exposure: Jiarui Xiong; Training of the classifier: Jialun Lin; Assessment of decoding accuracy for experimental and simulated data: Jialun Lin; Calculation of BSC channel capacity: Jiarui Xiong. Manuscript drafting: Jiarui Xiong (text section) and Jialun Lin (figure and movie section).  Manuscript revision and review: Fan Jin, Jun Chu, Jiarui Xiong, Jialun Lin, Liang Wang, Rongrong Zhang, Shuai Yang.
\section*{Declarations}

The authors declare no competing interests.
\bibliography{Ref}
\newpage
\begin{figure}[h]
	\centering
	\includegraphics[width=1.0\textwidth]{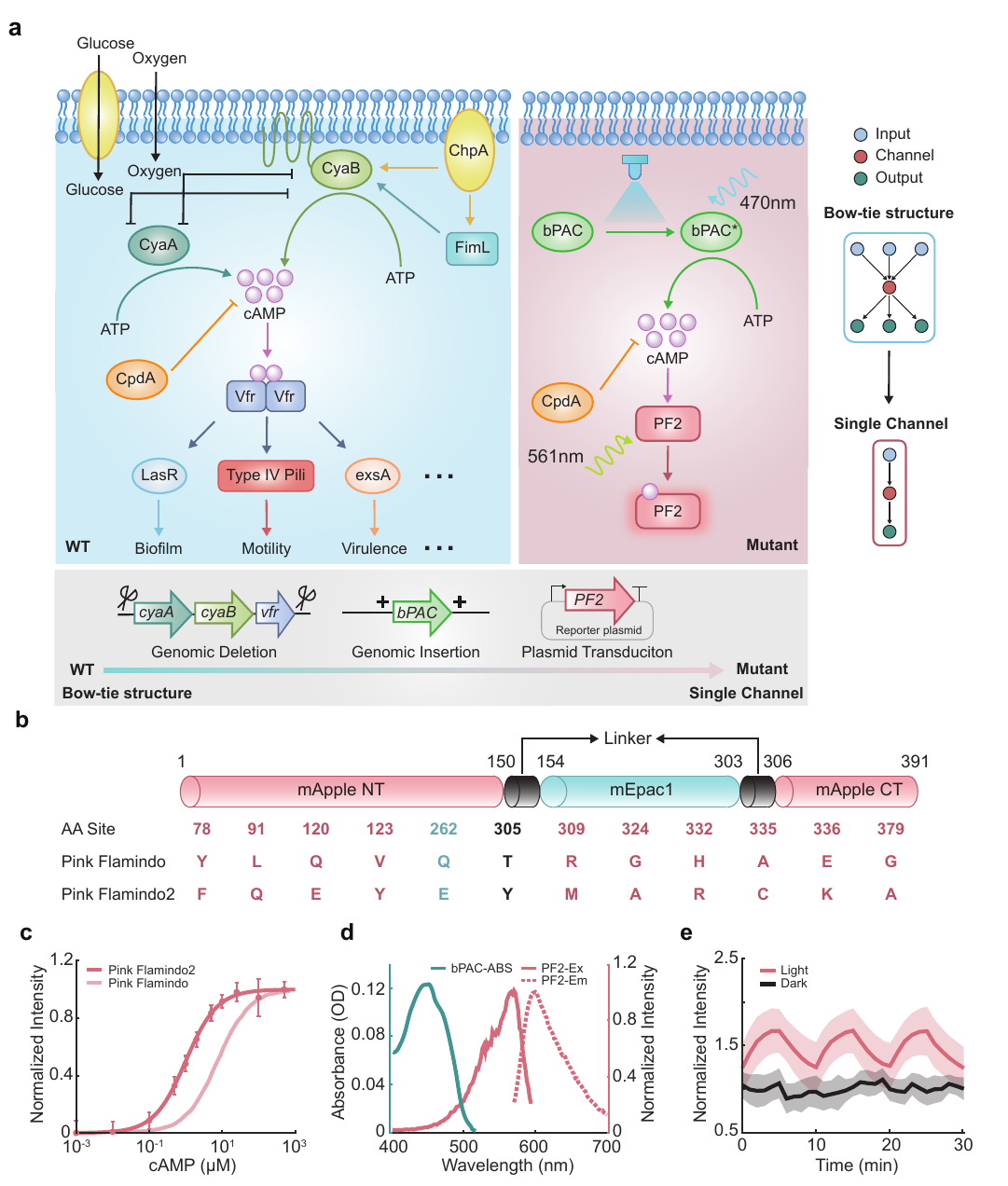}
	\raggedright
	\caption{\textbf{Engineering a synthetic circuit utilizing bPAC and PinkFlamindo2 for measuring information transmission via the cAMP pathway.} \
\normalfont (a) Experimental circuit design principles. The cAMP communication follows a bow-tie structure, where upstream signals are transmitted to the mid-layer through the synthesis of cAMP molecules catalyzed by adenylate cyclases CyaA and CyaB. Signals from the mid-layer to downstream are regulated globally upon cAMP binding to Vfr. By selectively knocking out key genes and introducing the optogenetic tool bPAC, along with cAMP molecular probe tools, a simplified quantitative system is constructed. This system utilizes light}
\label{fig1}		
\end{figure}
\clearpage
\begin{figure}[h]
	\raggedright
	\captionof*{figure}{\normalfont frequencies to assess the information transmission capability of the cAMP signaling pathway. (b) Schematic representation of the cAMP probe sequence. PinkFlamindo2 was developed by optimizing mutations of several amino acid residues from PinkFlamindo. (c) Dose-response curve of PinkFlamindo2 (red line) and PinkFlamindo (green line) to increasing concentrations of cAMP. The $K_d$ values calculated using the Hill equation were \SI{1.0}{\micro M} and \SI{7.2}{\micro M} for PinkFlamindo2 and PinkFlamindo, respectively. The data for PinkFlamindo2 were obtained from our  three independent experiments and are represented as the means ± standard deviation and the observed maximum fluorescence intensity was used for normalization. The data for PinkFlamindo were adopted from the literature.(d) Absorption spectra of purified bPAC in its light-adapted state (green line) and the excitation (solid red line) and emission spectra (dashed red line) of purified PinkFlamindo2 protein in the presence of cAMP. The disparity in peak positions between the absorption peak of bPAC and the excitation peak of PinkFlamindo2 indicates the decoupling of these two tools. The observed maximum fluorescence intensity was used for normalization. (e) Comparison of fluorescence oscillations in the strain co-expressing bPAC and PinkFlamindo2 under periodic illumination (with a 40-minute pre-treatment of periodic light) and those in the dark control, validating the decoupling of these two tools. The fluorescence intensity was normalized based on the mean value of the dark control group. The data represent the means ± standard deviation (1317 cells).}
\end{figure}
	
\begin{figure}[h]
	\centering
	\includegraphics[width=1.0\textwidth]{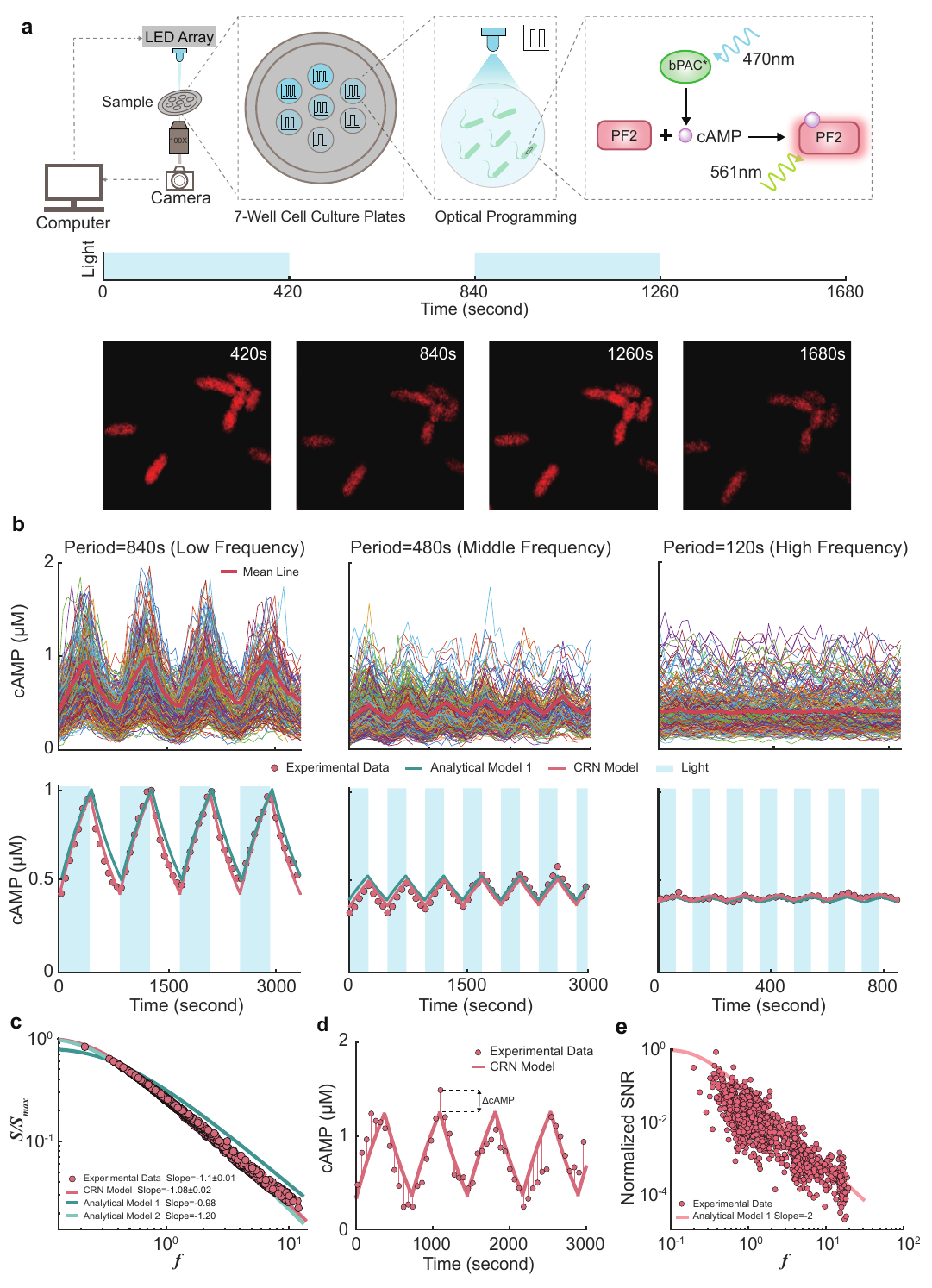}
	\raggedright
	\caption{\textbf{Quantitative analysis of cAMP signaling dynamics under modulated light stimulation at single-cell resolution.}\
\normalfont (a) Experimental setup and sequential images. A seven-well plate designed for the microscope apparatus was utilized to hold samples, accompanied by a corresponding} 
\label{fig2}
\end{figure}
\clearpage
\begin{figure}[h]
\raggedright
\captionof*{figure}{\normalfont LED light unit. Activation of bPAC was achieved using 471 nm wavelength light delivered under programmed control, while fluorescence excitation and imaging of PinkFlamindo2 were performed at 561 nm using computer-operated microscopy. The recorded images exhibited alternating bright and dark features under square wave illumination input modulated at a periodicity of 840 seconds.(b) Quantitative analysis of intracellular cAMP signaling dynamics at the single-cell level under illumination modulated at different frequencies. Measured cAMP oscillations within single bacteria are shown for low (840 s), medium (480 s), and high (120 s) frequency square-wave illumination inputs. The bold red lines represent the average values. By employing a Chemical Reaction Network (CRN) model, we successfully fitted the data and compared it with the conclusions derived from analytical model 1, demonstrating a high level of consistency between them.(c) The relationship between the normalized signal amplitude and the input frequency normalized by the cAMP degradation rate, based on empirical measurements (depicted as red dots), computational simulations (red solid line), and two separate theoretical predictions (dark green and light green solid lines). These data validate that in the high-frequency range, the signal exhibits an attenuation trend proportional to $f^{-1}$ when $f \gg1$.(d) The noise in cAMP fluctuations within individual bacteria was determined by fitting the experimental data (red dots) to a Chemical Reaction Network (CRN) model. The difference between the experimental data and the smooth curve obtained from the CRN model fitting (red line) represents the noise in cAMP oscillations within the single bacteria.(e) Relationship between the normalized signal-to-noise ratio (SNR) and normalized frequency. The theoretically derived curve (red line) aligns with the experimental data points (red dots). SNR exhibits an attenuation trend proportional to $f^{-2}$ when $f \gg1$.}
\end{figure}

\begin{figure}[h]
\centering
	
	\includegraphics[width=1.0\textwidth]{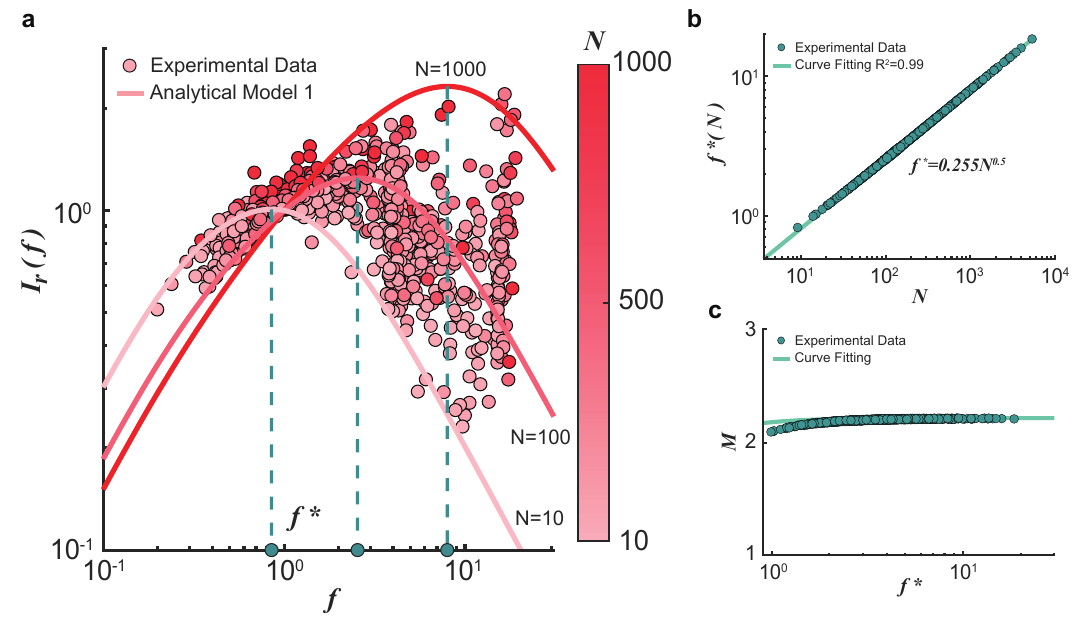}
	\raggedright
	\caption{\textbf{Relationships and characteristics in transmission rate, optimal frequency, and encoding scheme.} \
		\normalfont (a) Relationship between relative transmission rate ($Ir$) and normalized frequency ($f$) given molecular counts,  The theoretical curves (depicted as red lines)  pass through the experimental data points (represented by red dots). The color depth in the plot corresponds to the magnitude of molecular counts, with darker colors indicating higher counts. The concave downward shape of the curves signifies the presence of an optimal normalized frequency ($f^*$). (b) Relationship between the optimal normalized frequency ($f^*$) and cAMP molecular count ($N$) as determined from curve shapes. The empirical relationship $f^* = 0.255N^{0.5}$, obtained through fitting, exhibits an $R^2$ value greater than 0.99. This relationship demonstrates that the optimal normalized frequency increases with the molecular count, albeit at a decreasing rate of growth. (c) Relationship between the number of states ($M$) corresponding to the information at the optimal frequency and the optimal frequency ($f^*$). The variation in $f^*$ has minimal effect on $M$, which remains consistently stable around 2. This stability implies the adoption of a two-state encoding scheme.} 
\label{fig3}
\end{figure}

\begin{figure}
	\centering
	\includegraphics[width=1.0\textwidth]{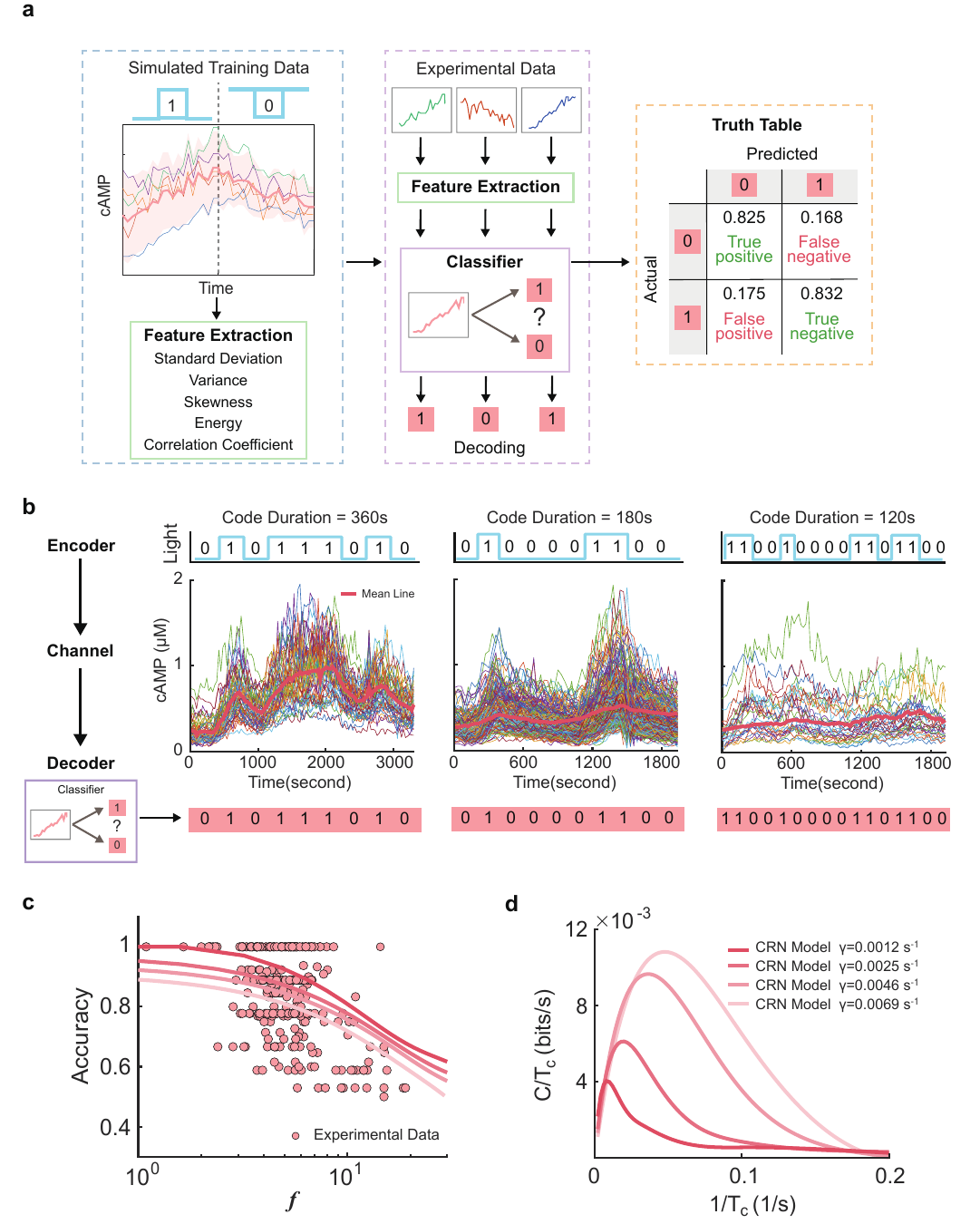}
	
	\raggedright
	\caption{\textbf{Decoding performance and information transmission of a cAMP communication system using machine learning.}  \
		\normalfont (a) The process of feature extraction from the simulated training dataset involved delineating fluctuation profiles of cAMP under illuminated (1) and non-illuminated (0) conditions. Subsequently, a network classifier model was developed using these extracted features.}
\label{fig4}
\end{figure}
\clearpage
\begin{figure}[h]
\caption*{\normalfont The enclosed truth table demonstrates the decoding framework's performance, where the trained neural network achieved an aggregated accuracy exceeding 80\% when applied to simulated data under varied code duration conditions. (b) The developed framework's decoding performance was assessed on experimental data comprising three distinct code duration settings of 360 s, 180 s, and 120 s. Applying the trained neural network model to these discrete test datasets, which featured varying code lengths, resulted in accurate decoding of the input states at the experimental end on average.(c) The relationship between decoding accuracy and reduction frequency is compared using experimental data from single bacterial samples (represented by red dots) and in silico data simulated using a CRN model (depicted by the red line).(d) The relationship between the information transmission rate and the code transfer frequency (1/Tc) is derived from the decoding accuracies achieved on simulated data. An optimal code duration exists where the rate is maximized. Distinct red lines representing varying apparent hydrolysis rates ($\gamma$) indicate that as $\gamma$ increases from low to high, the transmission rate increases from approximately 0.004 bits/s to 0.011 bits/s. Specifically, the analysis investigates how the information rate, calculated based on simulated decoding performance, changes with respect to the code frequency, identifying an optimal duration.}
\end{figure}

\end{document}